# High Mixing Entropy Enhanced Hydrogen Evolution Reaction in AlMnYNiCoAu Catalysts

Peng Zou[1,2], Bowen Zang[1,2], Lijian Song[1,2], Juntao Huo[1,2,*], Jun-Qiang Wang[1,2,*]

[1]*CAS Key Laboratory of Magnetic Materials and Devices, and Zhejiang Province Key Laboratory of Magnetic Materials and Application Technology, Ningbo Institute of Materials Technology and Engineering, Chinese Academy of Sciences, Ningbo 315201, China*

[2]*Center of Materials Science and Optoelectronics Engineering, University of Chinese Academy of Sciences, Beijing 100049, China*

An effective method for increasing the electrocatalytic activity of metallic glasses (MGs) in hydrogen evolution reaction (HER) is reported. This method applies a noble metal hybridization strategy to design a highly reactive catalyst for alkaline HER based on MGs with a nano-porous structure. The porous structure provides an abundance of active sites and exposes a large specific surface area to the electrolyte, thus effectively improving the activity of the HER catalysts. The doping of Au element can create electronic defects, adjust the electronic structure, change the electronic interaction, introduce lattice distortion, regulate mixing entropy, and improve the electrocatalytic performance of the catalyst. All these properties make np-AlMnYNiCoAu catalyst a potential electrode for alkaline HER. Significantly, we find that the high mixing entropy can enhance the HER performances. The present work could lead to a new approach to further develop environmentally friendly amorphous electrocatalyst with high efficiency and excellent stability for HER in alkaline solution.

* Correspondence should be addressed to:

J.T. Huo, huojuntao@nimte.ac.cn; J.Q. Wang, jqwang@nimte.ac.cn

## 1. Introduction

Sustainable hydrogen energy is a cornerstone of the future green economy.[1] Hydrogen evolution reactions (HER) through water splitting draws tremendous attention as a way for generating clean energy.[2] Pt-group metals are the most outstanding catalysts with low overpotential, small Tafel slope, and high exchange current density ($j_0$), whereas the low abundance and high cost limit their large-scale application in industry.[3] It is critical to investigate economical electrocatalysts with high activity and long stability for HER.[4] Moreover, the electrolysis cells and the electrodes often suffer corrosion in acidic solutions. An alkaline environment could not only mitigate the corrosion issue, but also let the earth-abundant transition metals be effectively utilized in HER.[5]

Over the past decades, in order to develop low-cost and efficient HER catalysts, researchers have developed several electrode materials such as non-noble transition metals[6-8] (e.g., Fe, Co, Ni and so on), phosphides,[9-11] nitrides,[12, 13] sulfides,[14-16] carbides,[17] and nanocarbon-based[18] materials. In addition, catalytic performance of metal-based electrodes also can be remarkably improved by alloying,[19, 20] forming nanostructures,[21, 22] doping,[23, 24] the existence of lattice defects,[25] and introducing strains.[26, 27] For example, reasonable doping of element can improve the charge transfer rate, increase the number density of active sites, and generate the synergy effect.[28] Xu et al. reported that NiFeP@N-CS shows low overpotential and good catalytic performance benefiting from the synergy effect and its hierarchical structure.[29] Furthermore, Liu et al. found that there are plenty of coordinatively unsaturated sites in higher energy states in a disordered atomic structure of IrNiTa metallic glass films. These features have been proved to be favorable for catalytic activity.[30] Moreover, J. Kruzic et al. prepared HER catalysts with nanocrystals in nanosponge-like high-entropy metallic glass (HEMG) ribbons through a facile dealloying process. The excellent HER performance of the de-alloyed HEMG originated from the synergistic functions of an enlarged surface area, self-constructed nanocrystals, and optimized electronic structures.[31]

One important challenge of designing advanced catalysts with lower cost is that the less noble elements addition often results in a poor electrocatalytic performance. The addition of multiple electrocatalytic elements could promote the synergy effect and hence achieve a good HER performance.[32, 33] In this study, we developed an $Al_{85-x}Mn_3Y_3Ni_6Co_3Au_x$ (x = 0 to 5) electrodes with a self-supporting nano-porous structure after de-alloying process. Ni, Co, and Au were added as active sites providers. Since Au is on the different side of the volcano plot to Ni and Co, it could bring the $\Delta G_{H^*}$ of the alloy closer to an intermediate value.[34, 35] We tuned the amount of Au doping to adjust the mixing entropy and investigate its influence on the HER performance of the de-alloyed $Al_{85-x}Mn_3Y_3Ni_6Co_3Au_x$ electrodes. Our results showed that instead of the $Al_{80}Mn_3Y_3Ni_6Co_3Au_5$ alloy, which had the largest amount of noble element, $Al_{82}Mn_3Y_3Ni_6Co_3Au_3$ alloy which has the highest mixing entropy after de-alloying showed the optimum HER performance in alkaline solution. An overpotential as low as 124 mV and a Tafel slope of 69 mV $dec^{-1}$ was achieved at a current density of 10 mA $cm^{-2}$. Moreover, benefiting from the introduction of Au, the nano-porous $Al_{82}Mn_3Y_3Ni_6Co_3Au_3$ electrocatalyst exhibited excellent durability with slight decay over 18 hours of continuous electrolyzing at 10 mA $cm^{-2}$ current density.

## 2. Experimental Methods

**2.1 Chemicals and Materials:** Aluminum pellets (99.99%, Φ6*6mm), manganese pellets (99.98%, 1-10mm), yttrium bulks (99.99%, 1-6mm), nickel pellets (99.9%, Φ3*3mm), cobalt bulks (99.95%, 1-10mm), gold wire (99.99%, Φ1.0mm), potassium hydroxide (GR, 85%) and absolute ethanol (AR, 99.8%) were purchased and used as received.

**2.2 Synthesis of Materials:** In this study, multi-component $Al_{82-x}Mn_3Y_3Ni_6Co_3Au_x$ metallic glasses were synthesized by the following method. The master alloys were prepared by melting the raw materials as listed above in an arc melting furnace under the protection of a Ti-gettered argon atmosphere. The master alloys were melted in a quartz tube by induction melting and then casted by a single-roller melt spinner in an argon atmosphere. The circumferential speed of the copper wheel used in melt spinning process was about 40 m s$^{-1}$. The as-spun ribbons were 1.5-2 mm in width and 20-30 μm in thickness.

**2.3 Preparation of Working Electrodes:** The as-spun ribbons were de-alloyed by immersing them in 1 M KOH solution for 10 mins. At next, the de-alloyed electrodes were activated by cyclic voltammetry. The corresponding actual compositions were characterized by energy-dispersive X-ray spectroscopy (EDS) in Table S1.

**2.4 Electrochemical Measurements:** The electrochemical measurements were carried out at room temperature using an electrochemical workstation (Zahner Zennium) with a three-electrode system. A Pt foil and an Ag/AgCl electrode (the concentration of Cl$^-$ in the electrode is 3.5M) were used as the counter electrode and the reference electrode, respectively. The electrolyte used in our electrochemical measurements was 1 M KOH aqueous solution. The nominal area of the samples immersed in the electrolyte for electrochemical tests was 20 mm$^2$. The HER linear sweep voltammetry (LSV) polarization curve was recorded at a scan rate of 10 mV s$^{-1}$. The stability test was performed using chronopotentiometry at a current density of -10 mA cm$^{-2}$ without iR compensation. The electrochemical impedance spectroscopy (EIS) was measured in the frequency range from 10 mHz to 100 kHz with an amplitude of 5 mV. To evaluate the electrochemical active surface area (ECSA) of the amorphous self-supporting electrode, the double-layer capacitance ($C_{dl}$) value was estimated by cyclic voltammetry (CV) in alkaline electrolyte according to the relationship ECSA = $C_{dl}/C_s$ with a $C_s$ value of 40 μF cm$^2$. After the electrochemical measurements, the ribbons were taken out from the electrolyte and washed by deionized water or anhydrous alcohol to neutrality. The overpotential was converted to the reversible hydrogen electrode (RHE) according to the following equation:

$$E_{RHE} = E_{Ag/AgCl} + 0.2046 + 0.059 \times pH$$

The preliminary electrochemical stability of all electrodes was investigated at 10 mA cm$^{-2}$ current density. The relevant electrochemical results were corrected by uncompensated resistance.

$$\eta_{HER} = E_{RHE} - E_{iR}$$

**2.5 Definition of Entropy:** The mixing entropy is mainly contributed by the configurational entropy. For the sake of simplicity, configurational entropy is a great substitute for mixing entropy. Then entropy can be calculated:[36]

$$\Delta S_{mix} = -R\left(\frac{1}{n}\ln\frac{1}{n} + \frac{1}{n}\ln\frac{1}{n} + \ldots + \frac{1}{n}\ln\frac{1}{n}\right) = -R\ln\frac{1}{n} = R\ln n$$

$$c_i = \frac{1}{n}$$

where R represents the gas constant, $c_i$ is the mole percentage of the $i^{th}$ component.

**2.6 Characterization Methods:** The amorphous structure of as-spun ribbons was confirmed by an X-ray diffraction (XRD, Bruker D8 Advance) using Cu K$_\alpha$ radiation at 40 kV and 40 mA. The microstructure of the de-alloyed ribbons was characterized by scanning electron microscopy (SEM, Hitachi S4800) and transmission electron microscopy (TEM, Tecnai F20). The TEM samples were prepared by focused ion beam (FIB, Auriga). Chemical composition analysis was performed by an energy dispersive X-ray spectrometer (EDS, Oxford Inca x-sight). The elemental valence was investigated using X-ray photoelectron spectrometer (XPS, Thermo ESCALAB 250XI) with a monochromatic Al K$_\alpha$ (1486.6 eV) source and a concentric hemispherical energy analyzer.

**3. Results and Discussion**

The as-pun $Al_{85-x}Mn_3Y_3Ni_6Co_3Au_x$ ribbons were fully amorphous proved by the XRD patterns in Supplementary Materials (Fig. S1). Figure 1a shows the normalized LSV curves of de-alloyed $Al_{85-x}Mn_3Y_3Ni_6Co_3Au_x$ electrodes in 1M KOH solution. The overpotential is plotted against the Au content in Fig. 2. The Tafel plots of the de-alloyed electrodes are given in Fig. 1b. The overpotential and the Tafel slope continuously decreases simultaneously as the Au content increases from 0 to 3 at%, which indicates that the Au doping up to 3 at% is beneficial to the electrocatalytic activity of the de-alloyed $Al_{85-x}Mn_3Y_3Ni_6Co_3Au_x$ electrodes. The exchange current density ($j_0$) of the de-alloyed electrodes estimated from the Tafel plots by extrapolation method (Fig. S2) reflects a promotive effect to HER kinetics by Au doping. The best HER performance among the $Al_{85-x}Mn_3Y_3Ni_6Co_3Au_x$ alloys studied is observed in the $Al_{82}Mn_3Y_3Ni_6Co_3Au_3$ (denoted as 'Au3') alloy, which shows the lowest overpotential (124 mV) and Tafel slope (69 mV dec$^{-1}$) at 10 mA cm$^{-2}$ current density and the fastest HER kinetics in Fig. S2. Further addition of Au beyond 3 at% deteriorates the HER performance, which is reflected by the rebound of overpotential and Tafel slope and the drop back of HER kinetics. The turnover frequency (TOF) of the $Al_{85-x}Mn_3Y_3Ni_6Co_3Au_x$ alloys developed in this work is higher than that of the non-noble metal electrocatalysts reported in other works as compared in Fig. 1d. In addition, they also have excellent electrochemical stability. Figure 1f shows that the HER performance of de-alloyed Au3 electrode only decays slightly (≈12 mV overpotential rise) over 18 h and 1000 cycles functioning.

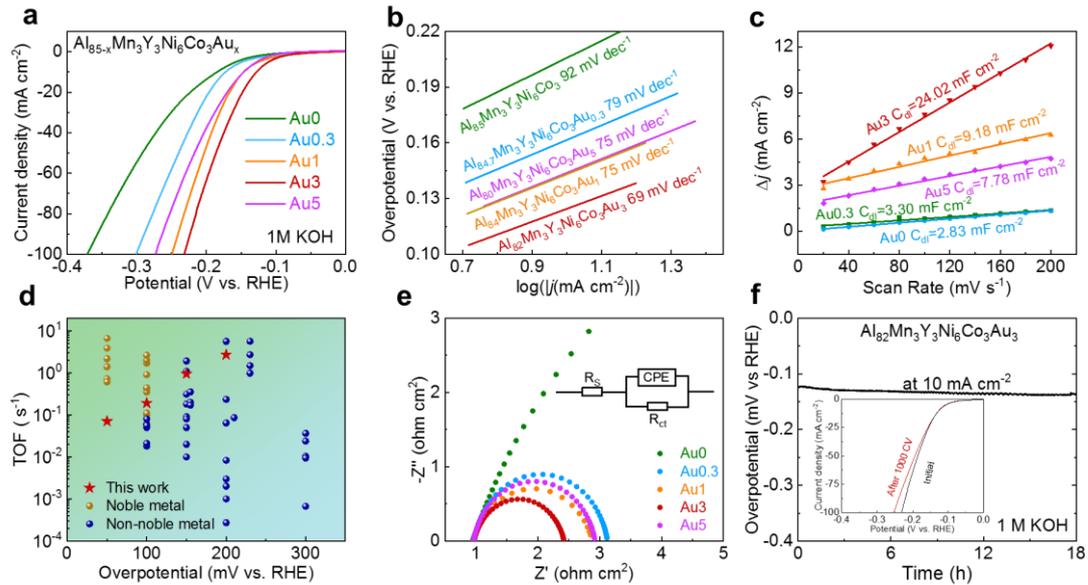

Fig. 1. (a) Linear sweep voltammetry (LSV) of catalysts recorded in 1 M KOH solution with iR-correction; (b) corresponding Tafel slope of the LSV curves in (a); (c) the capacitive current (estimated from the voltammetry scans in Fig. S3) as a function of scan rates; (d) the turnover frequency (TOF) at different overpotentials for various electrocatalysts in 1.0 M KOH solution; (e) Nyquist plot of nano-porous catalysts. The inset shows the equivalent circuit. (f) Chronopotentiometry stability test over 18 h at a constant current density of −10 mA cm$^{−2}$ (inset shows the initial polarization curve of the Au3 electrode and its polarization curve after 1000 cycles).

Figure 1c shows that the double layer capacitance of the de-alloyed electrodes with different alloy compositions. Since the electrochemical active surface area (ECSA) of an electrode is directly correlated to its double-layer capacitance,[37, 38] it could be realized from Fig. 1c that Au doping to $Al_{85-x}Mn_3Y_3Ni_6Co_3Au_x$ alloys increases the ECSA of the de-alloyed electrodes. This could be one of the reasons that Au doping improves the catalytic activity for HER. The charge-transfer resistance ($R_{ct}$) estimated from the Nyquist plots (Fig. 1e) is also reduced by Au doping. A small charge-transfer resistance represents a splendid electron pathway for electron transport and hence accelerated the HER kinetics[39, 40], which is another reason that Au doping improves the HER performance of de-alloyed $Al_{85-x}Mn_3Y_3Ni_6Co_3Au_x$ electrodes. The trend of ECSA and $R_{ct}$ with Au content is consistent with the trend of overpotential and Tafel slope as shown in Fig. 2.

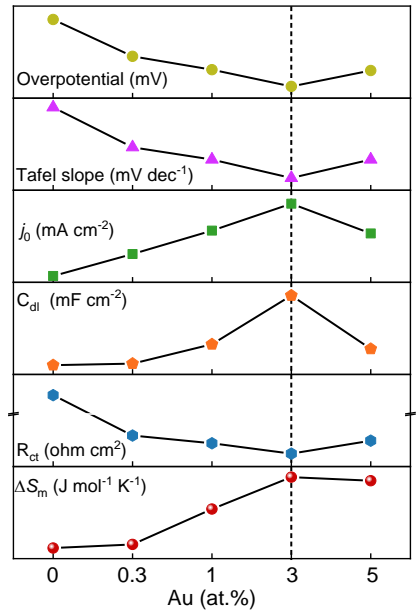

Fig. 2. Overpotential ($\eta$), Tafel slope, exchange current density ($j_0$), double-layer capacitances ($C_{dl}$), charge-transfer resistance ($R_{ct}$) and mixing entropy ($\Delta S_m$) as a function of Au content.

Interestingly, the mixing entropy of the de-alloyed electrodes (Table. S2) also shows the same trend as the HER performance in Fig. 2. Increasing the Au content initially increases the mixing entropy of the de alloyed electrodes, while excess amount of Au doping will reduce the mixing entropy. The change in mixing entropy influences the strength of synergy effect in the de-alloyed electrodes, which will impact the HER performance. Alloys with high mixing entropy can form new, tailorable active sites in multiple elements adjacent to each other. By rational selection of the structure and composition of the elements, the interactions can be tuned.[41] It has demonstrated that benefiting from the high entropy at nanoscale and strong synergy effects between active metals, high entropy alloys can have excellent catalytic performances and stability for HER.[42-44]

The increase in ESCA of the de-alloyed $Al_{85-x}Mn_3Y_3Ni_6Co_3Au_x$ electrodes by Au doping could be attributed to the microstructure change induced by Au addition. Figure 3a, 3c, and 3e shows the microstructure of the surface of de-alloyed $Al_{85-x}Mn_3Y_3Ni_6Co_3Au_x$ (x = 1, 3, and 5) electrodes. The SEM images indicate that the surface of all electrodes is in a nano-porous structure constructed by interconnected metallic ligaments. The elemental mapping of the Au3 sample (Fig. S4) proves that most of Al was removed from the alloy by de-alloying process and the rest constituent elements distributed homogeneously. The TEM images of the samples in Fig. 3a, 3c, and 3e are given by Fig. 3b, 3d, and 3f, respectively. Au is found to exist as nanocrystals (fcc Au, $d_{(111)}$ = 2.355 Å) in the de-alloyed electrodes. The SAED pattern of the Au3 sample in Fig. 3i also indicates the existence of fcc-Au in an amorphous matrix. These Au nanocrystals may inhibit the diffusion and rearrangement of atoms, as reported in nano-porous copper.[45] A decreasing pore size with the increasing Au content could be realized by comparing Fig. 3a, 3c, and 3e. The pore size of the Au3 sample shown in Fig. 3g is about 15 nm, which is consistent with that estimated from Fig. 3c. These nano-sized ligaments and pores provide efficient pathways for ion diffusion and a large specific surface area for active site exposure. For alloys with a similar Al content, reducing the pore size would increase the number density of pores and hence increase the specific surface area of de-alloyed electrodes. Thus, the ECSA of de-alloyed electrodes increases as the Au content increases until 3 at% of Au doping is achieved. Further reducing the pore size by increasing the Au content to 5 at% does not increase the ECSA further. This should be attributed to the too small pore size which could restrict the desorption of hydrogen generated in the HER process. As a result, a reduced exchange current density is observed in Fig. 1c for the Au5 sample, and the HER performance is deteriorated.

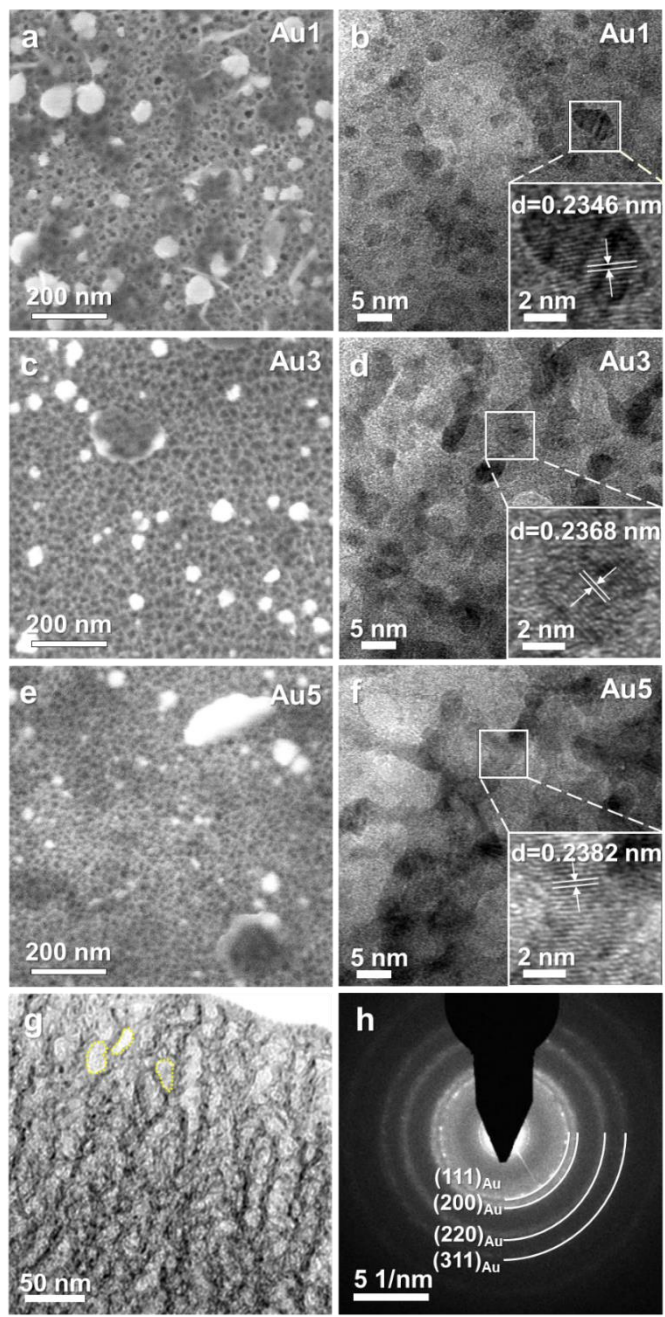

Fig. 3. (a), (c), and (e) SEM images of the surface of de-alloyed Au1, Au3 and Au5 ribbons. (b), (d), and (f) High-resolution TEM images of the de-alloyed Au1, Au3 and Au5 ribbons. (g) Cross-sectional TEM image of the de-alloyed Au3 ribbon. (h) SAED patterns of the de-alloyed Au3 ribbon.

Besides the reduction in pore size, the change in surface chemistry of de-alloyed $Al_{85-x}Mn_3Y_3Ni_6Co_3Au_x$ induced by Au doping also improves the HER performance. Fig. 4 shows the zoomed-in XPS spectra of de-alloyed $Al_{85-x}Mn_3Y_3Ni_6Co_3Au_x$ electrodes with different Au contents. Figure 4c and 4d shows the appearance of metallic Ni and Co in de-alloyed $Al_{85-x}Mn_3Y_3Ni_6Co_3Au_x$ electrodes with 1, 3, and 5 at% Au. The intensity of peaks corresponding to metallic Ni and Co increases has Au content increases. This indicates that the addition of Au could improve the corrosion resistance of amorphous $Al_{85-x}Mn_3Y_3Ni_6Co_3Au_x$ alloys. Metallic Ni and Co function more effectively than their oxidized state as active sites for HER. This becomes another origin that the addition of Au improves the HER performance of de-alloyed $Al_{85-x}Mn_3Y_3Ni_6Co_3Au_x$ electrodes. In addition, the presence of Au increases the amount of metal oxides and inhibits the formation of metal hydroxide during de-alloying process (Fig. 4f). The existence of metal hydroxide has been proved to be harmful to the active sites for HER.[5] Thus, Au doping would also improve the HER performance of amorphous $Al_{85-x}Mn_3Y_3Ni_6Co_x$ alloys via increasing the number density of active reaction sites.

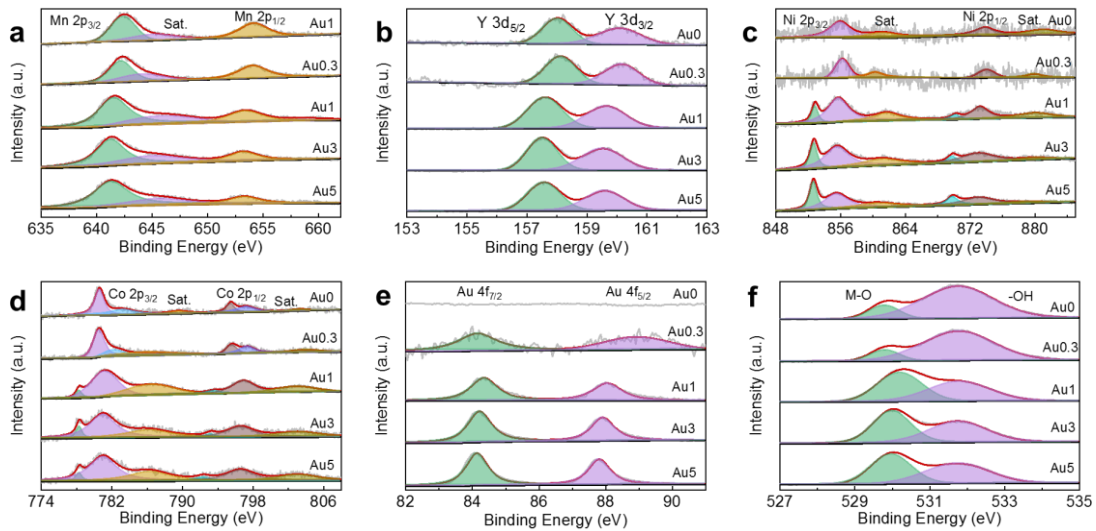

Fig. 4. High-resolution XPS spectra of (a) Mn 2p, (b) Y 3d, (c) Ni 2p, (d) Co 2p, (e) Au 4f, and (f) O 1s for the nano-porous electrode surfaces.

## 4. Conclusions

In summary, a facile synthetic route has been developed to produce Al-based amorphous alloys with self-supported nano-porous structure for efficient hydrogen evolution. Proper Au doping was found to be beneficial to the HER performance of the de-alloyed Al-based electrode by increasing the specific surface area and modifying the surface chemistry. The strong synergy effect of different elements in alloys with higher mixing entropy may also a reason of the improved HER performance. Optimum HER performance was observed in the alloy with 3 at% of Au and the highest mixing entropy. An overpotential of 124 mV and a Tafel slope of 69 mV dec$^{-1}$ were achieved by the de-alloyed Au3 alloy, which is much better than other non-noble electrocatalysts. Our research provides new ideas for designing efficient catalysts by adding multiple electrocatalytic elements and tuning the mixing entropy to optimize catalytic performance.

**Acknowledgments**

The authors acknowledge the financial supports from the National Key R&D Program of China (2018YFA0703604, 2018YFA0703602), National Natural Science Foundation of China (NSFC 51827801, 51922102, 52071327, 92163108), Youth Innovation Promotion Association CAS (No. 2019296), Zhejiang Provincial Natural Science Foundation of China (LR22E010004, 2022C01023).

# References

1. Y. Zhou, M. Niu, S. Zhu, Y. Liang, Z. Cui, X. Yang and A. Inoue, *Electrochimica Acta*, 2019, **296**, 397-406.
2. W. Xu, S. Zhu, Y. Liang, Z. Cui, X. Yang, A. Inoue and H. Wang, *Journal of Materials Chemistry A*, 2017, **5**, 18793-18800.
3. Q. T. Phan, K. C. Poon and H. Sato, *International Journal of Hydrogen Energy*, 2021, **46**, 14190-14211.
4. M. Y. Wu, P. F. Da, T. Zhang, J. Mao, H. Liu and T. Ling, *ACS Appl Mater Interfaces*, 2018, **10**, 17896-17902.
5. L. Li, P. Wang, Q. Shao and X. Huang, *Chemical Society Reviews*, 2020, **49**, 3072-3106.
6. S. E. Fosdick, S. P. Berglund, C. B. Mullins and R. M. Crooks, *ACS Catalysis*, 2014, **4**, 1332-1339.
7. P. Wang, K. Jiang, G. Wang, J. Yao and X. Huang, *Angew Chem Int Ed Engl*, 2016, **55**, 12859-12863.
8. Y. Yuan, J. Wang, S. Adimi, H. Shen, T. Thomas, R. Ma, J. P. Attfield and M. Yang, *Nat Mater*, 2019, **19**, 282-286.
9. F. H. Saadi, A. I. Carim, W. S. Drisdell, S. Gul, J. H. Baricuatro, J. Yano, M. P. Soriaga and N. S. Lewis, *J Am Chem Soc*, 2017, **139**, 12927-12930.
10. L. Feng, H. Vrubel, M. Bensimon and X. Hu, *Phys Chem Chem Phys*, 2014, **16**, 5917-5921.
11. Y. Tan, H. Wang, P. Liu, Y. Shen, C. Cheng, A. Hirata, T. Fujita, Z. Tang and M. Chen, *Energy & Environmental Science*, 2016, **9**, 2257-2261.
12. C. Lei, Y. Wang, Y. Hou, P. Liu, J. Yang, T. Zhang, X. Zhuang, M. Chen, B. Yang, L. Lei, C. Yuan, M. Qiu and X. Feng, *Energy & Environmental Science*, 2019, **12**, 149-156.
13. H. Yan, C. Tian, L. Wang, A. Wu, M. Meng, L. Zhao and H. Fu, *Angew Chem Int Ed Engl*, 2015, **54**, 6325-6329.
14. J. X. Feng, J. Q. Wu, Y. X. Tong and G. R. Li, *J Am Chem Soc*, 2018, **140**, 610-617.
15. Y. Guo, T. Park, J. W. Yi, J. Henzie, J. Kim, Z. Wang, B. Jiang, Y. Bando, Y. Sugahara, J. Tang and Y. Yamauchi, *Advanced Materials*, 2019, **31**, 1807134.
16. C. Tang, Z. Pu, Q. Liu, A. M. Asiri and X. Sun, *Electrochimica Acta*, 2015, **153**, 508-514.
17. Q. Gong, Y. Wang, Q. Hu, J. Zhou, R. Feng, P. N. Duchesne, P. Zhang, F. Chen, N. Han, Y. Li, C. Jin, Y. Li and S. T. Lee, *Nat Commun*, 2016, **7**, 13216.
18. H. Fan, H. Yu, Y. Zhang, Y. Zheng, Y. Luo, Z. Dai, B. Li, Y. Zong and Q. Yan, *Angew Chem Int Ed Engl*, 2017, **56**, 12566-12570.
19. X. Wang, R. Su, H. Aslan, J. Kibsgaard, S. Wendt, L. Meng, M. Dong, Y. Huang and F. Besenbacher, *Nano Energy*, 2015, **12**, 9-18.
20. S. Zhang, Y. Rui, X. Zhang, R. Sa, F. Zhou, R. Wang and X. Li, *Chemical Engineering Journal*, 2021, **417**, 128047.
21. X. B. Ge, L. Y. Chen, J. L. Kang, T. Fujita, A. Hirata, W. Zhang, J. H. Jiang and M. W. Chen, *Advanced Functional Materials*, 2013, **23**, 4156-4162.
22. Q. Wu, M. Luo, J. Han, W. Peng, Y. Zhao, D. Chen, M. Peng, J. Liu, F. M. F. De Groot and Y. Tan, *ACS Energy Letters*, 2020, **5**, 192-199.
23. C. Tang, R. Zhang, W. Lu, L. He, X. Jiang, A. M. Asiri and X. Sun, *Advanced Materials*, 2017, **29**, 1602441.

24. L. Yan, B. Zhang, J. L. Zhu, Y. Y. Li, P. Tsiakaras and P. K. Shen, *Appl. Catal. B-Environ.*, 2020, **265**, 11.
25. Y. Ma, M. Chen, H. Geng, H. Dong, P. Wu, X. Li, G. Guan and T. Wang, *Advanced Functional Materials*, 2020, **30**, 2000561.
26. K. Jiang, M. Luo, Z. Liu, M. Peng, D. Chen, Y.-R. Lu, T.-S. Chan, F. M. F. De Groot and Y. Tan, *Nature Communications*, 2021, **12**.
27. T. Ling, D.-Y. Yan, H. Wang, Y. Jiao, Z. Hu, Y. Zheng, L. Zheng, J. Mao, H. Liu, X.-W. Du, M. Jaroniec and S.-Z. Qiao, *Nature Communications*, 2017, **8**.
28. Y. Zhang, J. Xu, Y. G. Ding and C. D. Wang, *International Journal of Hydrogen Energy*, 2020, **45**, 17388-17397.
29. J. Hei, G. Xu, B. Wei, L. Zhang, H. Ding and D. Liu, *Applied Surface Science*, 2021, **549**, 149297.
30. Z. J. Wang, M. X. Li, J. H. Yu, X. B. Ge, Y. H. Liu and W. H. Wang, *Adv Mater*, 2019, DOI: 10.1002/adma.201906384, 1906384.
31. Z. Jia, K. Nomoto, Q. Wang, C. Kong, L. Sun, L. C. Zhang, S. X. Liang, J. Lu and J. J. Kruzic, *Advanced Functional Materials*, 2021, **31**, 2101586.
32. S. Ju, J. Feng, P. Zou, W. Xu, S. Wang, W. Gao, H.-J. Qiu, J. Huo and J.-Q. Wang, *Journal of Materials Chemistry A*, 2020, **8**, 3246-3251.
33. X. Liu, S. Ju, P. Zou, L. Song, W. Xu, J. Huo, J. Yi, G. Wang and J.-Q. Wang, *Journal of Alloys and Compounds*, 2021, DOI: 10.1016/j.jallcom.2021.160548, 160548.
34. J. Greeley, T. F. Jaramillo, J. Bonde, I. B. Chorkendorff and J. K. Norskov, *Nature Materials*, 2006, **5**, 909-913.
35. J. Kibsgaard, C. Tsai, K. Chan, J. D. Benck, J. K. Nørskov, F. Abild-Pedersen and T. F. Jaramillo, *Energy & Environmental Science*, 2015, **8**, 3022-3029.
36. Y. Zhang, D. Wang and S. Wang, *Small*, 2021, DOI: 10.1002/smll.202104339, 2104339.
37. Y. a. Zhu, Y. Pan, W. Dai and T. Lu, *ACS Applied Energy Materials*, 2020, **3**, 1319-1327.
38. T. Wang, X. Wang, Y. Liu, J. Zheng and X. Li, *Nano Energy*, 2016, **22**, 111-119.
39. C. Zhang, S. Liu, Z. Mao, X. Liang and B. Chen, *Journal of Materials Chemistry A*, 2017, **5**, 16646-16652.
40. X. Ma, Y. Zhu, S. Kim, Q. Liu, P. Byrley, Y. Wei, J. Zhang, K. Jiang, S. Fan, R. Yan and M. Liu, *Nano Letters*, 2016, **16**, 6896-6902.
41. H. Li, J. Lai, Z. Li and L. Wang, *Advanced Functional Materials*, 2021, **31**, 2106715.
42. M. Liu, Z. Zhang, F. Okejiri, S. Yang, S. Zhou and S. Dai, *Advanced Materials Interfaces*, 2019, **6**, 1900015.
43. T. Jin, X. Sang, R. R. Unocic, R. T. Kinch, X. Liu, J. Hu, H. Liu and S. Dai, *Advanced Materials*, 2018, **30**, 1707512.
44. J. W. Yeh, S. K. Chen, S. J. Lin, J. Y. Gan, T. S. Chin, T. T. Shun, C. H. Tsau and S. Y. Chang, *Adv. Eng. Mater.*, 2004, **6**, 299-303.
45. Z. Dan, F. Qin, Y. Sugawara, I. Muto and N. Hara, *MATERIALS TRANSACTIONS*, 2012, **53**, 1765-1769.

# Supplementary Material

# High Mixing Entropy Enhanced Hydrogen Evolution Reaction in AlMnYNiCoAu Catalysts

Peng Zou[1,2], Bowen Zang[1,2], Lijian Song[1,2], Juntao Huo[1,2,*], Jun-Qiang Wang[1,2,*]

[1]*CAS Key Laboratory of Magnetic Materials and Devices, and Zhejiang Province Key Laboratory of Magnetic Materials and Application Technology, Ningbo Institute of Materials Technology and Engineering, Chinese Academy of Sciences, Ningbo 315201, China*

[2]*Center of Materials Science and Optoelectronics Engineering, University of Chinese Academy of Sciences, Beijing 100049, China*

\* Correspondence should be addressed to:

J.T. Huo, huojuntao@nimte.ac.cn; J.Q. Wang, jqwang@nimte.ac.cn

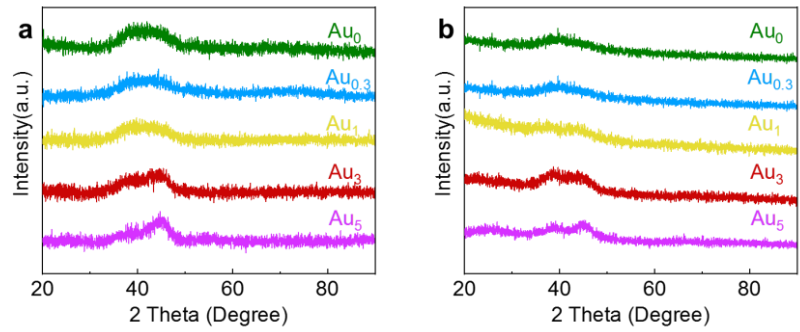

**Fig. S1** The XRD patterns for the (a) as-spun and (b) de-alloyed $Al_{85-x}Mn_3Y_3Ni_6Co_3Au_x$ (x = 0, 0.3, 1, 3, and 5) ribbons.

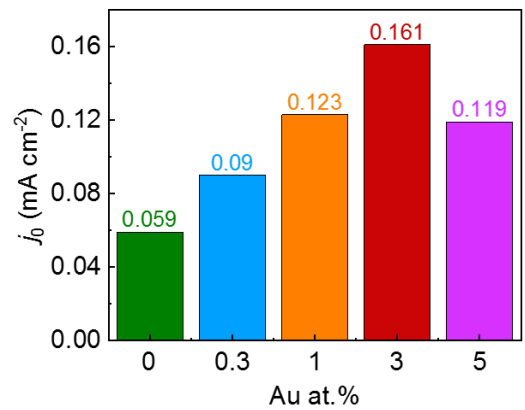

**Fig. S2** The exchange current density ($j_0$) of $Al_{85-x}Mn_3Y_3Ni_6Co_3Au_x$ (x = 0, 0.3, 1, 3, and 5) catalysts.

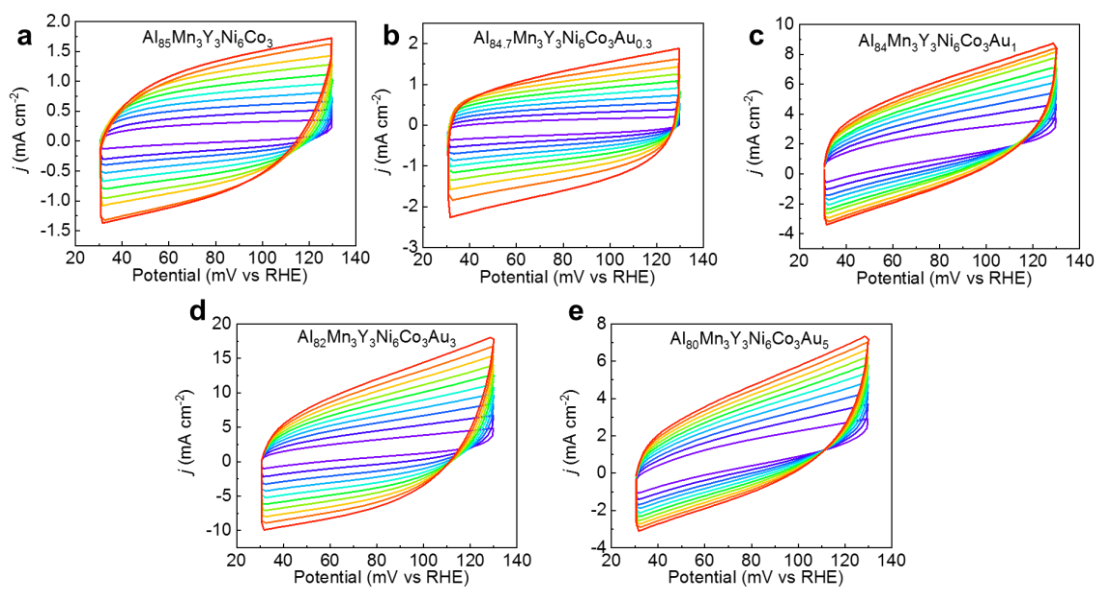

**Fig. S3** (a)-(e) The cyclic voltammograms (CVs) of nano-porous catalysts under 1.0 M KOH solution for $Al_{85-x}Mn_3Y_3Ni_6Co_3Au_x$ (x = 0, 0.3, 1, 3, and 5) ribbons.

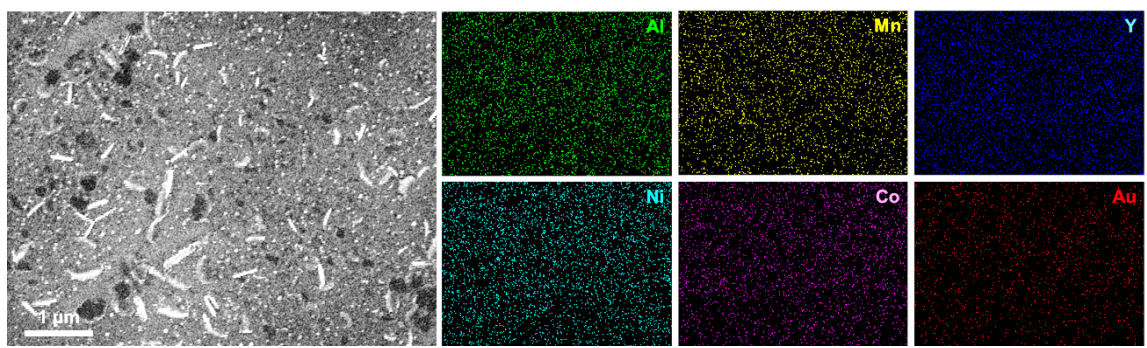

**Fig. S4** EDS elemental mapping images of Al, Mn, Y, Ni, Co and Au in de-alloyed $Al_{82}Mn_3Y_3Ni_6Co_3Au_3$ ribbon.

**Table S1.** The component determined from EDS for the de-alloyed samples, respectively.

| Elements (at. %) | Au0 | Au0.3 | Au1 | Au3 | Au5 |
|---|---|---|---|---|---|
| Al | 30.04 | 20.36 | 22.32 | 20.14 | 21.34 |
| Mn | 13.20 | 12.38 | 8.42 | 12.63 | 10.71 |
| Y | 16.33 | 25.21 | 15.71 | 12.77 | 11.43 |
| Ni | 26.54 | 27.98 | 31.64 | 26.91 | 22.89 |
| Co | 13.88 | 14.08 | 15.50 | 13.36 | 11.61 |
| Au | 0.00 | 0.00 | 6.42 | 14.19 | 22.01 |

**Table S2.** The estimated mixing entropy for the de-alloyed samples, respectively.

| | Au0 | Au0.3 | Au1 | Au3 | Au5 |
|---|---|---|---|---|---|
| $\Delta S_m$ (J/mol·K) | 1.55R | 1.56R | 1.66 R | 1.75 R | 1.74 R |